\documentclass[pre,preprint,showpacs]{revtex4}                  
             
\usepackage{graphicx}
\usepackage{amssymb}
\usepackage{amsmath}
\usepackage{subfigure,wrapfig}

\begin{document}

\title{Interface growth in two dimensions: A Loewner-equation approach}

\author{Miguel A.~Dur\'an} 
\author{Giovani L.~Vasconcelos}
\email{giovani@df.ufpe.br}

\affiliation{Laborat\'orio de F\'{\i}sica Te\'orica e Computacional, 
Departamento de F\'{\i}sica, Universidade Federal de Pernambuco,
50670-901, Recife, Brazil.}

\date{\today}

\begin{abstract}
The problem of  Laplacian growth  in two dimensions is considered within the  Loewner-equation framework. Initially the problem of fingered growth recently discussed by Gubiec and Szymczak [T.~Gubiec and P.~Szymczak, Phys.~Rev.~E {\bf 77}, 041602 (2008)] is revisited and a new exact solution for a three-finger configuration is reported. Then a general class of growth models for an interface growing in the upper-half plane is introduced  and the corresponding Loewner equation for the problem is derived. Several examples are given including interfaces with one or more tips  as well as multiple growing interfaces. A generalization of our interface growth model in terms of ``Loewner domains,'' where  the growth rule is specified by a time evolving measure, is briefly discussed.
\end{abstract}


\maketitle

\section{Introduction}

The Loewner equation \cite{loewner} 
is an important result in the theory of univalent functions \cite{univalent}  that has found  important applications  in nonlinear dynamics, statistical physics, and conformal field theory  \cite{review1, review2, BB, review3}.  In its most basic formulation, the Loewner equation  is a first-order differential equation for the conformal mapping $g_t(z)$ from a given ``physical domain,'' consisting of a complex region  $\mathbb{P}$  minus  a curve $\Gamma_t$ emanating  from its boundary, onto a ``mathematical domain'' represented by $\mathbb{P}$ itself. 
Usually, $\mathbb{P}$  is either the upper  half-plane or the exterior of the unit circle, but recently the Loewner equation for the channel geometry was also considered \cite{poloneses}.
The  Loewner equation depends on a driving function, here called $a(t)$, that is the image of the growing tip under the mapping $g_t(z)$. An important development on the theory of the Loewner equation  was the discovery by Schramm \cite{schramm} that when the driving function $a(t)$ is a Brownian motion the resulting Loewner evolution describes
 the scaling limit of certain statistical mechanics models. This result spurred great interest in the so-called stochastic Loewner equation \cite{review2,BB,review3}.

Recently, the deterministic Loewner equation  was used to study the problem of Laplacian fingered growth in both the half-plane and radial geometries \cite{makarov,selander} as well as in  the channel geometry \cite{poloneses}. In this class of models the growth takes place only at the tips of slit-like fingers and the driving function $a(t)$ has to follow a specific time evolution in order to ensure that the tip grows along gradient lines of the corresponding Laplacian field. In spite of its simplicity, the   model was able to reproduce some of the qualitative behavior seen in experiments on fingered growth \cite{poloneses,combustion}. Nevertheless,  treating the fingers as infinitesimally thin is a rather severe approximation and so this ``thin''  finger model is applicable only if the width of the fingers is much smaller than the separation between fingers. On the other hand, there are other instances of Laplacian growth where ``extended fingers'' (i.e., fingers with a finite, non-negligible  width) are observed  \cite{pelce} and for such cases a description in terms of Loewner evolutions is still lacking.

One of the main motivations of this paper is to seek to develop a framework based on the Loewner equation in which one can study more general growth problems in two dimensions. We start our analysis by first revisiting the problem of fingered growth discussed in \cite{poloneses} and report here a novel exact solution for the case of a symmetrical configuration with three fingers. We then move on to discuss a general class of growth models where an interface grows into the upper-half plane starting from a segment on the real axis. In our model, the growth rate is specified at a certain number of special points along the interface, referred to as ``tips'' and ``troughs'', which then determine the growth rate at the other points of the interface according to a specific growth rule (formulated in terms of a polygonal curve in the mathematical plane). By making appropriate use of the Schwarz-Christoffel transformation, 
we are then able to derive the Loewner equation for the problem. Several examples are given in which the interface evolution is numerically computed by a direct integration of the Loewner equation.  We also briefly discuss a more general formulation of the Loewner equation for interface growth in the upper-half plane where the growth rule is given in terms of a time evolving measure. This approach may open up the possibility for studying other interesting growth problems, such as the 
growth of fractal interfaces \cite{stanley}, within the context of stochastic Loewner evolutions.

The paper is organized as follows. In Sec.~\ref{sec:2} we briefly review the problem of fingered growth in the upper half-plane in the context of the chordal Loewner equation and report a new solution for a symmetrical configuration with three fingers. In Sec.~\ref{sec:3} we discuss a general class of growth models in which the  growing domain  is delimited by an interface in the upper half-plane. Here  we first derive in Sec.~\ref{sec:3a} a new Loewner equation for the case in which the growth rule is specified in terms of a polygonal curve in the mathematical plane. Several several examples are given in Sec.~\ref{sec:3b}, including the cases of interfaces with one or more tips as well as the case of multiple growing interfaces. We then briefly discuss in Sec.~\ref{sec:3c} a more general formulation for interface growth in the upper-half plane in terms of ``Loewner domains.''  Our main results and conclusions are summarized in Sec.~\ref{sec:4}.

\section{Fingered Growth and the Chordal Loewner Equation}
\label{sec:2}

\begin{figure}[t]
\begin{center}
\includegraphics[width=0.6\textwidth]{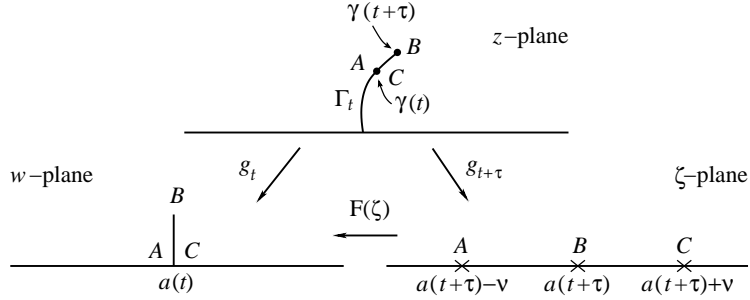}
\end{center}
\caption{The physical $z$-plane and the mathematical $w$- and  $\zeta$-planes at times $t$ and $t+\tau$, respectively, for a single  finger in the upper half-plane.   The mapping $g_t$ maps the curve $\Gamma_t$ onto a segment of the real axis on the $w$-plane, whereas the accrued portion of the curve during the infinitesimal time interval $\tau$ is mapped to a vertical slit. The mapping $g_{t}$ is obtained as the composition of $g_{t+\tau}$ and the slit mapping $F$; see text.}
\label{fig:1}
\end{figure}

In order to set the stage for the remainder of the paper and to establish the relevant notation, we begin our discussion by briefly reviewing  the problem of fingered growth in the context of the chordal Loewner equation. To this end we consider first the simplest  Loewner evolution, namely, that in which a curve starts from the real axis  at $t=0$ and then grows into the upper half-$z$-plane $\mathbb{H}$, where
\begin{equation*}
 \mathbb{H} = \{ z=x+iy \in \mathbb{C}: y > 0\}  .
\end{equation*}
The curve at time $t$ is denoted by $\Gamma_{t}$ and its growing tip is labeled by $\gamma(t)$.
Now let  $g_t(z)$ be the conformal mapping that maps the  ``physical domain,'' corresponding to the upper half-$z$-plane minus the curve $\Gamma_t$,  onto the upper half-plane of an auxiliary complex $w$-plane, called the ``mathematical plane,'' i.e., we have $w=g_t(z)$, where
 \begin{equation}
  g_t: \mathbb{H} \backslash \Gamma_{t} \rightarrow \mathbb{H}  ,
  \end{equation}
with the curve tip $\gamma(t)$ being mapped to a point $a(t)$ on the real axis in the $w$-plane, as shown in Fig.~\ref{fig:1}.  
Furthermore, we consider the growth process to be such that the accrued portion of the curve from $t$ to $t+\tau$, where $\tau$ is an infinitesimal time interval, is mapped under $g_{t}(z)$ to a vertical slit in the mathematical $w$-plane; see Fig.~\ref{fig:1}. The mapping function $g_t(z)$ must also satisfy the initial condition
\begin{equation}
g_{0}(z)=z,
\label{eq:g0}
\end{equation}
since we start with an empty  upper half-plane.
We also impose the so-called hydrodynamic normalization condition at infinity:
\begin{equation}
\label{tres}
 g_t(z) =z + O(\frac{1}{|z|}), \qquad {z \rightarrow \infty} .
\end{equation}
These conditions specify uniquely the mapping function $g_t(z)$.

From a more physical viewpoint, the problem formulated above belong to the class of Laplacian growth models where an interface evolves between two phases driven by 
a scalar field $\phi(x,y;t)$, representing, for example, temperature, pressure, or concentration, depending on the physical problem at hand.  In one phase, initially occupying the entire upper half-plane, the scalar field $\phi$ satisfies the Laplace equation
\begin{equation}
\nabla^{2}\phi=0,
\end{equation}
whereas in the other phase one considers $\phi$=const., say $\phi=0$, with the curve $\Gamma_t$ representing  a finger-like advancing interface between the two phases. (Here the finger is assumed to be infinitesimally thin.) The complex potential  for the problem can then be defined as $w(z,t)=\psi(x,y;t)+i\phi(x,y;t)$, where $\psi$ is the function harmonically conjugated to  $\phi$.
On the boundary of the physical domain, consisting here of the real axis together with the curve $\Gamma_t$, we impose the condition $\phi=0$, whereas at infinity we assume a uniform gradient field, $\vec{\nabla}\phi\approx \hat{y}$, or alternatively,
\begin{equation}
\label{eq:w}
w(z,t) \approx z, \qquad {z \rightarrow \infty} .
\end{equation}
From this point of view, the mapping function $g_{t}(z)$ introduced above corresponds precisely to the complex potential $w(z,t)$ of the problem. In particular, the fact that  in the $w$-plane the curve grows along a vertical line implies that the finger tip grows along gradient lines in the $z$-plane. To specify completely a given physical model, one has also to prescribe the interface velocity, which is usually taken to be proportional to some power $\eta$ of the gradient field: 
\begin{equation}
v_n\sim|\vec{\nabla}\phi|^{\eta}.
\label{eq:vn}
\end{equation}
We anticipate that for a single growing finger the specific velocity model  is not relevant in the sense that the finger shape will be independent of  the exponent $\eta$, which only affects the time scale of the problem. 
(For multifingers, however, different $\eta$'s may yield different patterns  \cite{poloneses}.)

For convenience of notation, we shall represent  the mathematical plane at time $t+\tau$ as the complex $\zeta$-plane 
and so we write $\zeta=g_{t+\tau}(z)$.  Now consider  the mapping
$w=F(\zeta)$, from 
 the upper half-$\zeta$-plane onto the mathematical domain in the $w$-plane; see Fig.~\ref{fig:1}.  The mapping function $g_{t+\tau}(z)$ can then be given in terms of $g_{t}(z)$ by 
\begin{equation}
 g_{t+\tau}=F^{-1}\circ g_{t},
 \label{eq:1f}
\end{equation}
 where $F^{-1}$ is the inverse of $F(\zeta)$.  The above relation  governs the time evolution of the function $g_t(z)$ and naturally  leads to the Loewner equation. A standard way  of showing this is to construct the slit mapping
$F(\zeta)$ explicitly, substitute its inverse in (\ref{eq:1f}), and then take the limit $\tau\to0$. One then finds the so-called {\it chordal Loewner equation}:
\begin{equation}
 \dot{g}_t(z) = \frac{d(t)}{g_t(z)-a(t)},
\label{loewner} 
\end{equation}
together with the condition
 \begin{equation}
\dot{a}(t)=0,
\label{eq:dota}
\end{equation}
so that $a(t)=a_0=$ const., which implies that the tip $\gamma(t)=g_t^{-1}(a(t))$ simply traces out a vertical line in the $z$-plane. 
%Alternatively, for the inverse mapping $f_{t}(w)=g_{t}^{-1}(w)$ we have
%\begin{equation}
% \dot{f}_t(w) = \frac{d(t)}{g_t(z)-a(t)},
%\label{loewner} 
%\end{equation}
The growth factor
 $d(t)$  is related to the tip velocity 
 by \cite{poloneses}
\begin{equation}
d(t)=|f_{t}^{\prime\prime}(a(t))|^{-\eta/2-1},
\label{eq:dteta}
\end{equation}
 where $f_{t}(w)$ is the inverse of $g_{t}(z)$. As already anticipated, 
 we can  take $d(t)=d_{0}$ (which corresponds to $\eta=-2$) by  reparametrizing the time coordinate.
 %The usual choice   for historical reasons is $d_{0}=2$  \cite{BB}. 
 
 In the case of multiple growing fingers \cite{poloneses}, the chordal Loewner equation becomes  \begin{equation}
 \dot{g}_t = \sum_{i=1}^{n} \frac{d_i(t)}{g_t-{a}_i(t)}, 
\label{eq:11}
\end{equation}
with the time evolution of the singularities $a_i(t)$ given by 
\begin{equation}
 \dot{a}_i(t) = \sum_{ \stackrel{j=1}{j\ne i}}^n \frac{d_j(t)}{{a}_i(t)-{a}_j(t)} .
 \label{eq:12}
\end{equation}
We recall that conditions (\ref{eq:12}) ensure that the fingers grow along gradient lines, meaning that the path traced by each finger tip $\gamma_{i}(t)$ from time $t$ to $t+\tau$ is mapped by $g_{t}(z)$ onto a vertical slit. Notice also that
if the growth factors $d_{i}(t)$ are all the same they can be assumed to be constant, say,  $d_{i}(t)=1$ (which corresponds to $\eta=-2$), since this amounts to a mere rescaling of the time variable. 
%(Note however that in the case of competing fingers we can fix only one of the $d_{i}$'s.)
For two symmetrical fingers, i.e.,  $d_1=d_2$,  equation (\ref{eq:11}) can be integrated exactly to yield the mapping function $g_t(z)$, from which the finger shapes can be computed analytically \cite{poloneses}. A related exact solution for two fingers was reported in \cite{kadanoff2}. 

\begin{figure}[t]
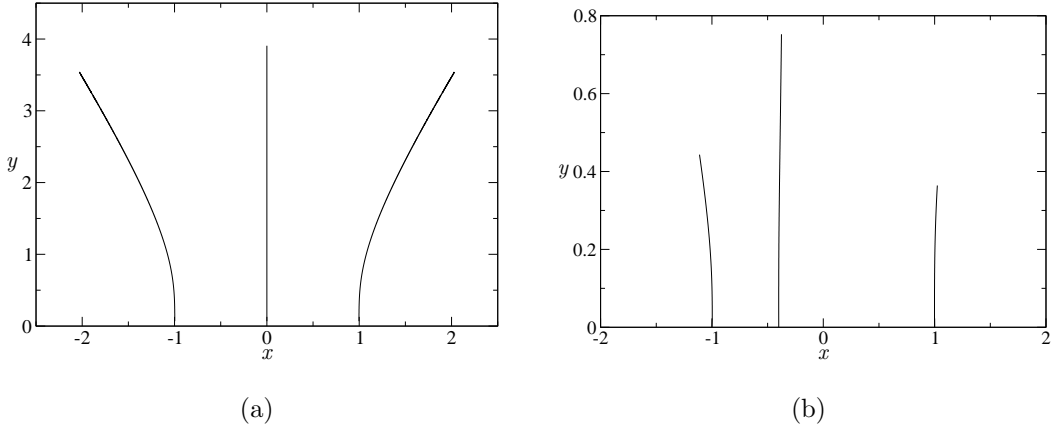

\begin{center}
\subfigure[\label{fig:2a}]{\includegraphics[width=0.4\textwidth]{fig2a.eps} }\hspace{5mm}
\subfigure[\label{fig:2b}]{\includegraphics[width=0.4\textwidth]{fig2b.eps} }
\end{center}
\caption{ Three fingers growing in the upper-half plane. In (a) we have a symmetrical configuration with $d_1=d_2=d_2=1$ and initial positions of the fingers at $a_3(0)=-a_1(0)=1$ and $a_2(0)=0$; in (b) we have  $d_1=1$, $d_2=5$, $d_3=0.9$, while the  initial positions are $a_3(0)=-a_1(0)=1$ and $a_2(0)=-0.4$.}
\end{figure}

An exact solution for $g_{t}(z)$ 
can also be obtained for a symmetrical configuration with three fingers where the middle finger grows vertically, say, along the $y$ axis, and the two flanking fingers are the mirror images of one another with respect to the $y$ axis. This situation is attained by choosing identical growth factors, $d_1=d_2=d_{3}=d_{0}$, and symmetrical initial conditions, $a_{3}(0)=-a_{1}(0)=a_0$ and $a_{2}(0)=0$. This initial symmetry is, of course, preserved by the dynamics and so we have $a_{2}(t)=0$ and $a_{3}(t)=-a_{1}(t)=a({t})$ for all $t\ge0$, which implies from (\ref{eq:12}) that 
\begin{equation}
a(t)=\sqrt{a_{0}^2+3d_{0}t}.
\end{equation}
 Using  this result one can integrate exactly the Loewner equation (\ref{eq:11}) and  obtain, after a straightforward calculation, the positions  $\gamma_{1,3}(t)$ of the tips of the  flanking fingers in terms of the following implicit equation
\begin{equation}
[\gamma_{1,3}(t)^2-\beta_{+}]^{1+\alpha}[\gamma_{1,3}(t)^2-\beta_{-}]^{1-\alpha}=(1-\beta_{+})^{1+\alpha}(1-\beta_{-})^{1-\alpha}(a_{0}^2+3d_{0}t)^{2},
\label{eq:3f}
\end{equation}
where
%$\alpha= {3}/\sqrt{57}$ 
$\alpha=\sqrt{3/19}$ 
and
$
\beta_{\pm}=(9\pm\sqrt{57})/{6}
$. In (\ref{eq:3f}) one  considers only the roots that lie in the upper-half plane and satisfy the initial condition $\gamma_{1,3}=\mp a_0$. In Fig.~\ref{fig:2a} we plot this solution for $a_{0}=1$ and $d_{0}=1$. 
From this figure one can see  that the side fingers are `repelled' by the middle finger and tend to straight lines  for sufficiently large times.  In fact, one can verify from (\ref{eq:3f})  that in the asymptotic limit $t\to\infty$ the flanking fingers approach the straight lines
%arg$(z)=\frac{\pi}{4}(1+3/\sqrt{57})$ 
arg$(z)=\frac{\pi}{4}(1+\sqrt{3/19})$ and arg$(z)=\frac{\pi}{4}(3-\sqrt{3/19})$.
%arg$(z)=\frac{3\pi}{4}(1-1/\sqrt{57})$.

If the growth factors are kept all equal but the initial condition is no longer symmetric, then the middle finger will initially be more  repelled by the closest finger,  thus leading to a symmetrical configuration in the limit $t\to\infty$ where the  asymptotic (vertical) position of the middle finger now depends on the initial condition.   On the other hand,  if the growth factors are not all the same then the finger with greatest $d_i$ will grow faster and screen the others. Here, however, the screening is partial in the sense that the ratio between the velocity of a slower finger and that of the fastest finger reaches a positive constant (different from unity).  An example of this case is shown  in Fig.~\ref{fig:2b} for the early stages of a three-finger evolution. In order to generate the curves shown in this figure we used the numerical scheme described in \cite{kadanoff2}. More specifically, we  start with the ``terminal condition'' $g_t=a_i(t)$ and integrate the Loewner equation (\ref{eq:11}) backwards in time, using a Runge-Kutta method of second order, to get the position of the respective tip $\gamma_i(t)=g_0$. We dealt with the pole singularities in (\ref{eq:11}) in a manner similar to that used in   \cite{kadanoff2}, which in our case meant finding the appropriate root of a fourth-degree polynomial that gives the value of $g$ at time $t-dt$, from where the Loewner can be integrated backwards to yield $g_0$.  
It is interesting to notice that the curves in Fig.~\ref{fig:2b} resemble certain fingering patterns seen in combustion experiments, see, e.g., Fig.~1e of Ref.~\cite{combustion}, even though  the solutions above are valid for the upper-half plane, whereas the experiments take place in a channel geometry, and we have used constant  growth factors (although with different values for each finger) rather than 
equation (\ref{eq:dteta}).

Before leaving this section we wish to  recall here that the evolution entailed by the chordal Loewner equation, as given by (\ref{loewner}) or   (\ref{eq:11}), describes a rather general class of  ``local growth'' processes  \cite{BB} in which there is one curve  or a set of curves growing from the real axis into the upper-half plane $\mathbb{H}$. 
Indeed, it is a theorem  \cite{review3,BB} that if  $\Gamma_{[0,\infty]}$  is a simple curve from the origin to $\infty$ in the upper-half plane $\mathbb{H}$,  then there exists a function $a(t)$ such that the curve is generated by a  Loewner evolution.  Conversely, if $a(t)$ is a sufficiently well-behaved function (more specifically, H\"older continuous with exponent  $> 1/2$), then $\gamma_t=g_t^{-1}(a(t))$ is a growing curve in $\mathbb{H}$. More general growth processes not restricted to growing curves can be generated by the so-called ``Loewner chains" \cite{BB}:
\begin{equation}
\dot{g}_t(z)=\int_{\mathbb{R}}\frac{\rho(x) dx}{g_{t}(z)-x}\, ,
\label{eq:Lc}
\end{equation}
where the density of singularities $\rho(x)$ can be viewed as a measure of the growth rate at a point $z$ at the boundary of the growing set that is the preimage of $x$ under $g_{t}(z)$. Since the shape of the growing domain is fully encoded in the map $g_{t}(z)$, equation (\ref{eq:Lc}) specifies the growth model once the density $\rho(x)$ is known. 
%Nonlinear dynamics can then arise if $\rho(x)$ is expressed in terms of $g_{t}(z)$. 
Although the formalism of Loewner chains is rather general, its practical usefulness is somewhat limited by the requirement that  the relevant density $\rho(x)$ must be specified {\it a priori}. In 
this context,  it should be noted that the only specific Loewner evolution  that has been studied in detail concerns the local growth models mentioned above where the density is a sum of Dirac $\delta$ functions \cite{BB}. In the next section, we will study a new class of  interface growth models where the growth velocity is specified at certain points of the interface, which then determine the growth rate at all the other points according to a specific rule.

\section{Interface Growth in the Half-Plane}
\label{sec:3}

\subsection{Generalized Loewner equation for a growing interface}
\label{sec:3a}

\begin{figure}[t]
\centering
\includegraphics[width=0.8\textwidth]{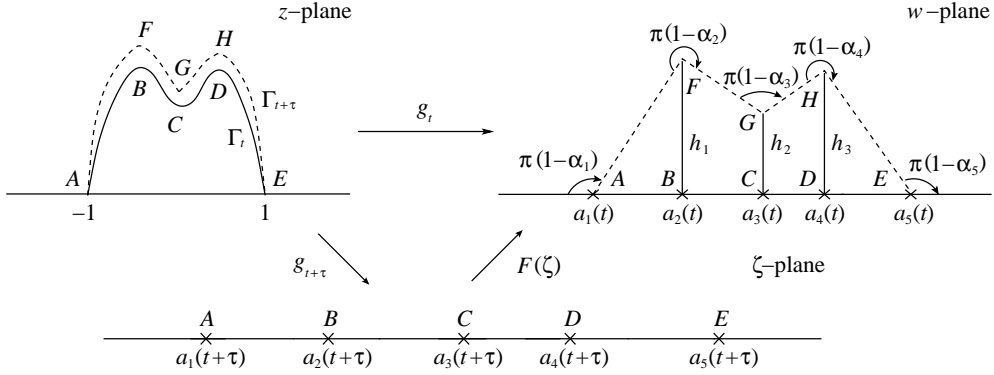}
\caption{The physical and mathematical planes for a growing interface. The mapping $g_{t}(z)$ maps the interface $\Gamma_t$ to a segment on the real axis, while the interface  $\Gamma_{t+\tau}$ is mapped
to a polygonal curve; see text for details.}
\label{fig:5}
\end{figure}

Here we consider the problem of an interface starting initially from a segment, say, $[-1,1]$,  along the real axis and growing into the upper half-$z$-plane, as indicated in Fig.~\ref{fig:5}.  
We suppose that the growing interface has a 
certain number of special points, referred to as {\it tips}   
%(points B and C in Fig.~\ref{fig:5}) 
or  {\it troughs}, where the growth rate is a local maximum or a local minimum, respectively, while the interface endpoints remain fixed, as illustrated in Fig.~\ref{fig:5} for the case of an interface with two tips (points B and D) and one trough (point C). 
Let us now denote by $\Gamma_t$ the interface at time $t$  and by $K_t$ the growing region delimited by $\Gamma_t$ and the real axis. Here we assume that the curve $\Gamma_t$ is simple so that the domain $D_t=\mathbb{H}\backslash K_t$ is simply connected. (In more technical terms, $K_t$ is a hull \cite{BB}.) We then consider the mapping $w=g_t(z)$ from the physical  domain $D_t$ in the $z$-plane to the upper-half plane $\mathbb{H}$  in the mathematical $w$-plane:
\begin{equation*}
 g_t:D_t \rightarrow \mathbb{H}  ,
\end{equation*}
where the mapping function $g_t(z)$ is required to satisfy the hydrodynamic normalization condition
\begin{equation}
g_t(z) =z + O(\frac{1}{|z|}), \qquad {z \rightarrow \infty},
\label{eq:gz}
\end{equation}
together with the initial condition 
\begin{equation}
g_0(z)=z.
\end{equation}
By definition, the interface $\Gamma_t$ is mapped under $g_t(z)$ to an interval $[a_1(t),a_N(t)]$ on the real axis of the $w$-plane, where $a_1(t)$ and $a_N(t)$ are the images of the endpoints $z=\pm1$, while the tips and troughs are mapped to the points $a_i(t)$, $i=2,...,N-1$, with $N$ being the total number of special points (i.e., tips, troughs and endpoints) of the interface. For instance, in Fig.~\ref{fig:5} we have $N=5$.

The growth dynamics is specified by requiring that the tips and troughs of the interface grow along gradient lines, while the  interface endpoints  remain ``pinned''   at $z=\pm 1$, in such a way that  the interface $\Gamma_{t+\tau}$ at time $t+\tau$, for infinitesimal $\tau$,  is mapped under $g_t(z)$ to a polygonal  curve in the $w$-plane,  as shown in  Fig.~\ref{fig:5}. 
The domain $D_{t+\tau}=\mathbb{H}\backslash K_{t+\tau}$ is mapped under $g_t(z)$ to   a degenerate polygon whose interior angle at the $i$-th vertex  is denoted by $\pi(1-\alpha_{i})$, with the convention that if the angle is greater than $\pi$ the corresponding parameter $\alpha_{i}$ is negative. It is easy to verify that the parameters $\alpha_{i}$'s satisfy the following relation
\begin{equation}
\sum_{i=1}^{N}\alpha_i=0,
\label{eq:sumd0}
\end{equation}
where we recall $N$ is the number of vertices of the polygonal curve in the $w$-plane.
If we now denote by   $h_{i}$ the height of the $i$-th vertex as measured from the real axis (see  Fig.~\ref{fig:5}), it is clear that  $h_{i}$  and hence the angle parameters $\alpha_{i}$ all go to zero as $\tau \to 0$. 
%In fact, we will see below that  $\alpha_{i}$ are linearly related  to  $\tau$ in the limit  $\tau \to 0$. 
In this limit one obtains another relation among the $\alpha_{i}$'s: 
 \begin{equation}
\sum_{i=1}^{N}a_{i} \alpha_i =0.
\label{eq:sd20}
\end{equation}

Let us know derive the Loewner evolution for the growth process described above. Since the domain in the $w$-plane has a polygonal shape, the mapping $w=F(\zeta)$ can be obtained from the Schwarz-Christoffel transformation \cite{CKP}:
\begin{equation}
 g_t = F(g_{t+\tau})= \int_{\zeta_{0}}^{g_{t+\tau}} \prod_{i=1}^N{[\zeta-a_i(t+\tau)]^{-\alpha_i}}\,d\zeta + F(\zeta_{0}),
 \label{eq:gtB}
\end{equation}
where $\zeta_0$ is a point to be chosen appropriately; see below. 
The integral  in (\ref{eq:gtB}) cannot be performed exactly for arbitrary $\alpha_{i}$'s, and so in order to obtain the Loewner equation for this case we need an alternative approach. First we expand the integrand in (\ref{eq:gtB}) up to first order in
the infinitesimal parameters $\alpha_i$'s, thus getting
\begin{equation}
 g_t \approx \int_{\zeta_{0}}^{g_{t+\tau}} \left\{ 1 - \sum_{i=1}^N{\alpha_i}\ln[\zeta-a_i(t+\tau)]\right\} d\zeta + F(\zeta_{0}),
\end{equation}
which after integration becomes
\begin{equation}
g_t=g_{t+\tau} -\zeta_{0}-G(g_{t+\tau})+G(\zeta_{0})  + F(\zeta_{0}) ,
\label{eq:1b}
\end{equation}
where
\begin{equation}
G(\zeta)= \sum_{i=1}^N \alpha_i[\zeta -a_i(t+\tau) ]\ln[\zeta -a_i(t+\tau)] 
\label{eq:1c}
\end{equation}
Now expanding  (\ref{eq:1b}) up to first order in $\tau$,
%\begin{equation}
% \dot{g}_t(z)=-\frac{G(g_{t+\tau})}{\tau} +\frac{G(\zeta_{0})  - F(\zeta_{0})+\zeta_{0}}{\tau} ,
%\label{eq:1b}
%\end{equation}  
taking $\tau \to 0$, and using the boundary condition  $\lim_{g_{t} \to \infty}\dot{g}_t=0$, which follows from (\ref{tres}),  one then obtains 
the following generalized Loewner equation
\begin{equation}
 \dot{g}_t(z)= \sum_{i=1}^N d_i(t)[g_{t} -a_i(t) ]\ln[g_{t} -a_i(t)] ,
\label{eq:1}
\end{equation}
together with the generic condition
\begin{equation}
\lim_{\tau\to0} \ \frac{\zeta_{0}-F(\zeta_{0})-G(\zeta_{0})}{\tau}=0,
\label{eq:2a}
\end{equation}
where the growth factors $d_i(t)$ appearing in (\ref{eq:1}) are defined by
\begin{equation}
d_i(t)= \lim_{\tau \to 0}\frac{\alpha_i}{\tau}.
\label{eq:2}
\end{equation}

The equations governing  the time evolution for the functions $a_i(t)$ can now be obtained by
considering $\zeta_0=a_i(t+\tau)$ in (\ref{eq:2a}), for $i=1,...,N$, with $F(\zeta_{0})$ then being the respective image of $a_{i}(t+\tau)$ in the $w$-plane:
\begin{equation}
F(a_i(t+\tau)) = a_i(t) +ih_{i}, 
\label{eq:Fz0}
\end{equation}
where for the endpoints we have $h_{1}=h_{N}=0$. 
Using (\ref{eq:Fz0}) in (\ref{eq:2a}) and performing a straightforward calculation, one finds 
\begin{equation}
 \dot{a}_i= \sum_{\stackrel{j=1}{j\ne i}}^N d_j(t)(a_i -a_j)\ln|a_{i} -a_j|.
\label{eq:ai}
\end{equation}

In view of  (\ref{eq:2}), equations (\ref{eq:sumd0}) and (\ref{eq:sd20}) for the parameters $\alpha_i$ imply equivalent relations for the growth factors $d_i(t)$:
\begin{equation}
\sum_{i=1}^{N}d_i=0,
\label{eq:sumd}
\end{equation}
 \begin{equation}
\sum_{i=1}^{N}a_{i} d_i =0.
\label{eq:sd2}
\end{equation}
From these equations
we can then express the parameter functions, $d_{1}(t)$ and $d_{N}(t)$, of the endpoints  in terms of the other functions $d_{i}(t)$, so that the growth model
 described by the Loewner evolution (\ref{eq:1}) and (\ref{eq:ai}) is completely specified by prescribing the growth factors $d_{i}(t)$ for the tips and troughs of the interface. [We remark parenthetically that although our growth model naturally allows for vertices that correspond to neither tips nor troughs we shall not consider such cases here; see however Sec.~\ref{sec:3c} for a discussion about more general growth rules.] The parameter functions $d_{i}(t)$ could in principle be related to the velocities of the corresponding points on the interface by  equations analogous to (\ref{eq:dteta}). For instance, in the simplest case $\eta=-2$ we have $d_{i}(t)=d_{0}$.  For general $\eta$, however,  the function $d_{i}(t)$ has a complicated implicit dependence on the function $g_{t}(z)$  involving the second derivative of its inverse, which renders the numerical integration of the Loewner equation (\ref{eq:1}) very difficult  (if possible at all). On the same token, the alternative approach for  computing Loewner evolutions based on direct iteration of slit mappings  \cite{poloneses, kennedy2}     is unlikely to result very useful in the case of growing interfaces, because the relevant mapping 
  to be iterated is not known in closed form; see (\ref{eq:gtB}). For these reasons, we shall consider in the examples below only  the case  $d_{i}(t)=d_{i}={\rm const.}$ (although not all $d_{i}$'s need  be the same), for which  the Loewner equation (\ref{eq:1}) can be easily integrated. 
%We expect nevertheless that the qualitative behavior  seen in this simpler situation  should also be valid for more general cases.
 
\begin{figure}[t]
\centering
\includegraphics[width=0.6\textwidth]{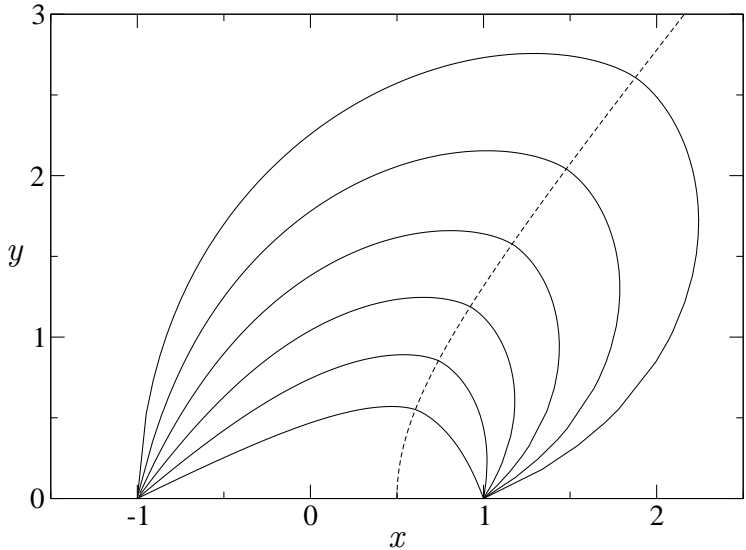}
\caption{Loewner evolution for a non-symmetric interface growing in the upper half-plane. The solid curves show the interface  at various times $t$, starting from $t=0.5$ up to $t=2.0$, with  a time separation of  $\Delta t =0.3$ between successive curves, whereas the dashed line indicates the  tip trajectory.}
\label{fig:4}
\end{figure}

\subsection{Examples}
\label{sec:3b}

We start by considering the case in which the interface has only one tip so that the number of vertices is obviously $N=3$. We can then solve  (\ref{eq:sumd}) and (\ref{eq:sd2}) to obtain the growth factors $d_{1}(t)$ and $d_{3}(t)$ as a function of $d_{2}(t)$:
\begin{equation}
 d_1(t) = - \frac{a_3(t)-a_2(t)}{a_3(t)-a_1(t)}d_2(t), \qquad d_3(t) = - \frac{a_2(t)-a_1(t)}{a_3(t)-a_1(t)}d_2(t).
 \label{eq:di}
\end{equation}
As mentioned above, the growth factor $d_{2}(t)$ can in principle be related to the tip velocity  but  here the specific form of $d_{2}(t)$  is not relevant for the shape evolution, for it only defines the time parameter, and so we set $d_2(t)=-1$. (Recall that according to our convention the growth factors for tips are negative while those for troughs are positive.) As for the initial conditions, we have  $a_1(0)=-1$ and $a_3(0)=1$,  corresponding to endpoints  $z=\pm1$, so that we have to specify only the initial location, $a_2(0)$,  of the tip. If we start with the tip at the center, i.e., $a_{2}(0)=0$, we obtain a symmetrically growing interface \cite{us}, since in this case equations  (\ref{eq:ai}) imply that $a_{2}(t)=0$ and $a_{1}(t)=-a_{3}(t)$ for all $t$. 
If, on the other hand, we start with the tip off-center,  i.e., $a_2(0)\ne0$, we then get an asymmetric growing interface (which can be regarded as an ``extended finger'' in the sense that it encloses a nonzero area). An example of such case for $a_2(0)=0.5$ is shown in Fig.~\ref{fig:4}, where the solid curves represent the interface  at various times $t$ starting from $t=0.5$ up to $t=2.0$ with  a time separation of  $\Delta t =0.3$ between successive curves, while the dashed curve  represents the path traced by the tip $\gamma_{t}=g^{-1}_{t}(a_{2}(t))$.  As one can see in this figure, for longer times the tip moves along a straight line in the $z$-plane,
%(of slope approximately 4/3 in this particular case), 
indicating that the initial asymmetry persists for all times.  [In this and subsequent figures the trajectories of the tips and troughs are computed, for clarity, up to a final time that is slightly greater than the final time for the complete  interfaces.] To generate the curves shown in Fig.~\ref{fig:4} we used a numerical scheme similar to that described in Sec.~\ref{sec:2} for the case of multifingers, namely, we integrate the Loewner equation (\ref{eq:1}) backwards in time with   terminal conditions $g_t=w$, for $w\in (a_{1}(t),a_{3}(t))$, to get the corresponding points $z=g_0$ on the interface.   
%Notice that here there is now singularity at the endpoints

\begin{figure}[t]
\centering
\includegraphics[width=0.6\textwidth]{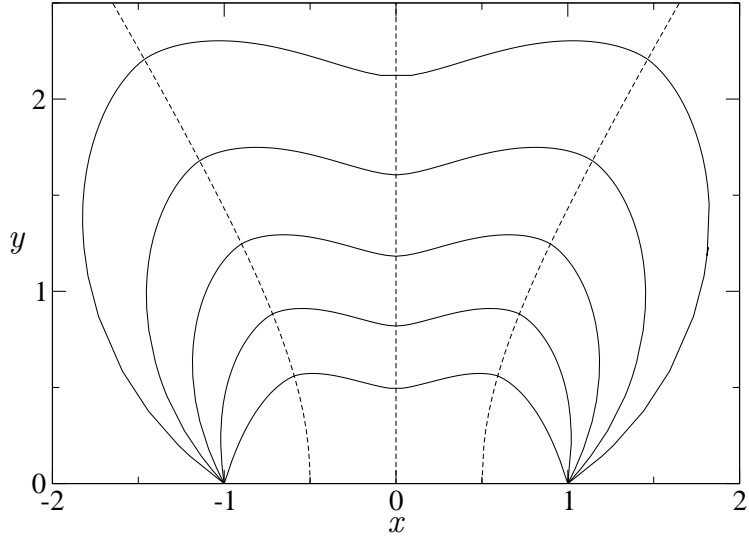}
\caption{Symmetric growing interface with two tips. Here $d_{2}=d_{4}=-1$ and $d_3=0.5$.
The solid curves represent the interface at times from $t=0.5$ to $t=1.7$,
with $\Delta t = 0.3$ between sucessive curves, whereas the dashed lines indicate the trajectories of  the tips and the trough.
}
\label{fig:6}
\end{figure}

Next we consider the case in which the interface has two tips and one trough. In this case we have $N=5$, and from (\ref{eq:sumd}) and (\ref{eq:sd2}) one can easily write the parameters $d_{1}$ and $d_{5}$ in terms of the growth factors $d_{2}$ and $d_{4}$ for the tips and $d_{3}$ for the trough. In Fig.~\ref{fig:6} we show an example of a symmetrical growing interface (solid curves),
% at times varying from $t=0.5$ to $t=1.7$,with a time interval $\Delta t = 0.3$ between sucessive curves, 
where the dashed lines represent the trajectories of the tips and the trough.
In this figure,  we started with symmetrical initial conditions, namely, $a_{5}(0)=-a_{1}(0)=1$, $a_{4}(0)=-a_{2}(0)=0.5$, and $a_{3}(0)=0$, and chose the growth factors of the two tips to be the same, $d_{2}=d_{4}=-1$, so as to preserve the initial symmetry, with $d_3=0.5$.  Note that the paths traced by the two tips and the trough are somewhat reminiscent of the trajectories observed in
the case of three symmetrical fingers discussed in Sec.~\ref{sec:2}; see Fig.~\ref{fig:2a}. 
On the other hand, if we start with a symmetric initial condition but one of the tips has a larger growth factor, then the fastest tip will move ahead 
of the other tip thus breaking the initial symmetry. For $t\to\infty$ the slower tip and the trough will both tend to merge  with the closest endpoint resulting in a  growing interface with only one tip. An example of this case is shown in Fig.~\ref{fig:6}
where  the tendency of the trough and the second tip to  merge is clearly evident. The behavior seen in this figure mimics somewhat the competition between initial protuberances in the early stages of the fingering instability \cite{pelce}. (In our model the competition is of course encoded in the choice of the growth factors.)

\begin{figure}[t]
\centering
\includegraphics[width=0.6\textwidth]{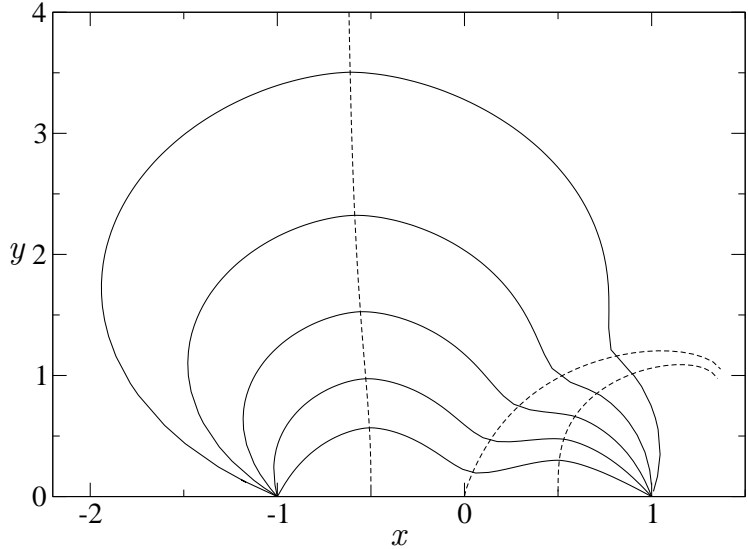}
\caption{Asymmetric growing interface with two tips. The initial conditions are the same as in Fig.~\ref{fig:6} but here we have $d_{2}=-1$, $d_{3}=0.8$, and $d_{4}=-0.5$. }
\label{fig:6b}
\end{figure}

The generic Loewner equation (\ref{eq:1}) can also describe the problem of multiple  growing interfaces.
% starting from disjoint intervals of the real axis. 
 In this case each interface $\Gamma^i_t$, for $i=1,...,n$, where $n$ is the number of distinct interfaces,  will be mapped under $g_t(z)$ to a corresponding interval on the real axis in the $w$-plane. Similarly, each advanced interface  $\Gamma^i_{t+\tau}$ is mapped by $g_t(z)$ to a polygonal curve in the $w$-plane. One can then readily convince oneself that the generic Loewner evolution defined by (\ref{eq:1}) and (\ref{eq:ai}) applies to this case as well, where  $N$ is now the total number of vertices corresponding to the sum of the number of vertices of each interface. 
%In what follows we discuss some examples for the case  in which there are two growing interfaces with one tip each.

%\vspace{1.0cm}

\begin{figure}[t]
\centering
\includegraphics[width=0.6\textwidth]{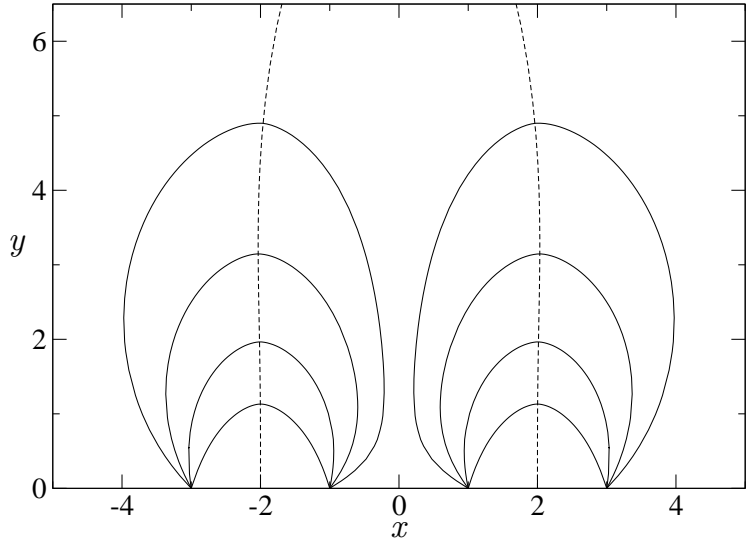}
\caption{Loewner evolution for two symmetrical growing interfaces. The interfaces start respectively at $[-3,-1]$ and $[1,3]$,
and the growth factor is $|d|=1$ for each tip. The solid curves represent the interfaces at times varying from $t=0.5$ up $t=1.4$,  with time intervals of $\Delta t = 0.3$, whereas the dashed line indicate the trajectories of the tips.}
\label{fig:8}
\end{figure}

In Fig.~\ref{fig:8} we show numerical solutions for the case of two symmetrical  interfaces
with one tip each. Since each interface has three special points (two endpoints and one tip)  we then have $N=6$.
% where the interfaces are the mirror images of one another with respect to the $y$ axis. 
Here the interfaces start to grow from the intervals $[-3,-1]$ and $[1,3]$, respectively, with the corresponding tips starting at symmetrical points $a_4(0)=-a_2(0)=2$ and having the same growth factors. 
%[Recall that the other parameter functions $d_{i}(t)$ can be obtained from $d_{1}$ and $d_{4}$ from the relations analogous to (\ref{eq:di}) for each interface.] 
The curves shown in Fig.~\ref{fig:8} were calculated for $d_{2}=d_{4}=-1$, at times varying from $t=0.5$ up $t=1.4$ with a time interval $\Delta t =0.3$ between successive curves. Notice that, as time goes by, the inner sides of the two interfaces  move towards one another leaving a narrow channel between them. For sufficiently large time, the width of such a channel becomes infinitesimally small so that for all practical purposes the resulting evolution will look like a single  symmetrical growing interface. An evidence of this fact is clearly seen in Fig.~\ref{fig:8} where the two tips (dashed lines) tend to ``attract'' each other and will eventually ``merge.'' It is worth mentioning, however, that it is hard to integrate the Loewner equation past the latest time shown in Fig.~\ref{fig:8}, because the functions $a_{3}(t)$ and $a_{4}(t)$ become essentially identical (within the computer resolution), rendering the numerical integration  very difficult after this point.

\begin{figure}[t]
\centering
\includegraphics[width=0.6\textwidth]{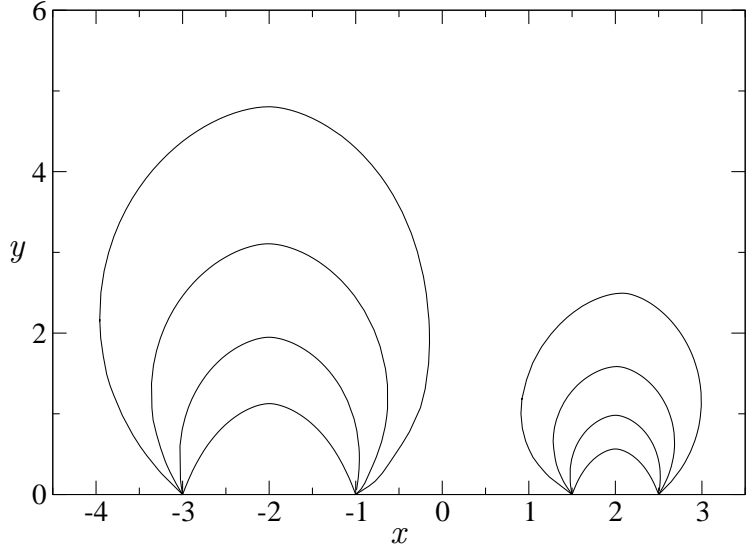}
\caption{Loewner evolution for two asymmetric growing surfaces. Here 
the growth factors and the initial width of first interface are the same as in
Fig.~\ref{fig:8} but the second interface starts to grow from the  narrower interval $[1.5,2.5]$. The different curves are for 
the same times as in  Fig.~\ref{fig:8}.}
\label{fig:10}
\end{figure}

In Fig.~\ref{fig:10} we show for comparison the case in which the two tips have identical growth factors but where  the interfaces now have different initial widths. We can see from this figure that the wider interface grows faster than the narrower one. This competition between the two growing interfaces resembles the so-called ``shadowing effect" in fingering phenomena, whereby the longer fingers grow faster and hinder the growth of the shorter fingers in their vicinities \cite{poloneses}.  [A similar effect is obtained if we start with interfaces of equal width but with one of the tips having a larger growth factor than the other one.]  
We have thus seen that in spite of its simplifying assumptions, chiefly among them the unbounded geometry and the polygonal growth rule, our model exhibits certain dynamical features such as finger competition and screening that are qualitatively similar  to what is  commonly observed in Laplacian growth processes  \cite{pelce}. We will show next that it is possible, in principle, to  extend the model to include  more general growth rules that may describe more accurately  the dynamics of specific growth processes.
%In this context, it is worth pointing out  that the polygonal growth rule described in Fig.~\ref{fig:5} ensures only that tips and troughs grow along gradient lines and does not necessarily enforce that the normal velocity at other points of the interface is proportional to the gradient of the scalar field, as in usual Laplacian growth.   

\subsection{Loewner Domains}
\label{sec:3c}

Here we consider from a more formal viewpoint  the Loewner evolution of a family of increasing hulls $K_t$ in the upper-half plane $\mathbb{H}$.
%, where $K_t\subset K_{t+s}$ for any $s>0$. 
We denote by $\Gamma_{t}=K_{t}\, \cap\, \mathbb{H}$ the growing interface at time $t$ and consider the mapping $g_{t}:\mathbb{H}\backslash K_{t} \to \mathbb{H}$ that
maps  $\Gamma_{t}$  to the interval $[a(t),b(t)]$ on the real axis. To specify the growth rule let us consider the infinitesimal hull $K_{\tau,t}=g(K_{t+\tau}\backslash K_t)$ in the $w$-plane, where  $\tau$ is an infinitesimal time step.  Note that in the growth models of  Sec.~\ref{sec:3a} the hull  $K_{t,\tau}$ corresponds to the interior of the polygon defined by $g_{t}(\Gamma_{t+\tau})$ and the real axis; see Fig.~\ref{fig:5}. Here we  consider more general growth problems in which  $K_{t,\tau}$ 
%of  $K_{t,\tau}$ in $\mathbb{H}$ 
is not necessarily bounded  by a piecewise-linear curve in $\mathbb{H}$. 
To be more precise, we consider the infinitesimal hull $K_{\tau,t}=\{w=x+i\tau y, 0< y < h_{t}(x)\}$ made of 
the set of points included between the real axis and the curve $y=\tau h_{t}(x)$, $\tau\ll 1$, with $x$  real. 
%[Here for convenience of notation we have written $w=x+iy$.]  
We assume that   $h_{t}(x)$ is a simple curve that is at least  twice differentiable. 
Let us now denote by $a^*_{\tau}(t)\le a(t)$ and  $b^*_{\tau}(t)\ge b(t)$ the endpoints of the curve $h_{t}(x)$, that is, $h_{t}(a^*_{\tau}(t))=h_{t}(b^*_{\tau}(t))=0$, where $\lim_{\tau\to0} a^*_{\tau}(t)=a(t)$ and $\lim_{\tau\to0}b^*_{\tau}(t)=b(t)$. Note that in the general case one has   $a^*_{\tau}(t)\ne a(t)$ and $b^*_{\tau}(t)\ne b(t)$, which allows for the endpoints $z_{1}(t)$ and $z_{2}(t)$ of the growing interface $\Gamma_{t}$ to move as time goes by. Thus, in contrast with the model discussed earlier, here the endpoints are not necessarily kept fixed.

Let us now introduce  a partition of the interval $[a^*_{\tau}(t), b^*_{\tau}(t)]$: $a^*_{\tau}(t)=a_{1}(t)<a_{2}(t)<\cdots<a_{N}(t)=b^*_{\tau}(t)$, and let $\tilde{h}_t(x)$ the piecewise-linear approximation of $h_{t}(x)$ defined by this partition:  $\tilde{h}_t(a_i)=h(a_i)$, with
$\tilde{h}_t(x)$, for $x\in[a_i(t),a_{i+1}(t)]$, being given by a straight line.   It is clear that the evolution entailed by the polygonal curve $\tilde{h}_t(x)$ is described in terms of the Loewner equation (\ref{eq:1}). If we now take the limit $N\to\infty$, we then get the following evolution equation:
\begin{equation}
\dot{g}_{t}(z)=\int_{a(t)}^{b(t)} \kappa_{t}(x)[g_{t}(z)-x]\ln[g_{t}(z)-x]dx, 
\label{eq:LD}
\end{equation}
where 
\begin{equation}
\kappa_{t}(x)=h_{t}^{\prime\prime}(x),
%\frac{d^2 h_{t}}{dx^2}.
\label{eq:rho}
\end{equation}
%and
%\begin{equation}
%h^\prime_{t}(a^*_{\tau}(t))=h_{t}^\prime(b^*_{\tau}(t))=0,
%\end{equation}
%as implied by conditions (\ref{eq:sumd}) and (\ref{eq:sd2}). 
with prime denoting derivative with respect to the argument $x$. 
Since  the shape of the growing domain is fully encoded in the map $g_{t}(z)$, equation (\ref{eq:LD}) specifies the growth model once the density $\kappa_t(x)$ is known. Note that 
the class of interface growth models described in Sec.~\ref{sec:3a} corresponds to the case when the density is a finite sum of Dirac $\delta$ peaks.

More generally, if we introduce a signed measure $\mu_{t}(x)$ satisfying the conditions
 \begin{equation}
\int_{\mathbb{R}}d\mu_{t}(x)=0,
\end{equation}
\begin{equation}
\int_{\mathbb{R}} x d\mu_{t}(x)=0,
\end{equation}
which are the analog of (\ref{eq:sumd}) and (\ref{eq:sd2}), 
we can then define the Loewner evolution generated by this measure by the following equation
\begin{equation}
\dot{g}_{t}(z)=\int_{\mathbb{R}} [g_{t}(z)-x]\ln[g_{t}(z)-x] d\mu_{t}(x) .
\label{eq:LDmu}
\end{equation}
Here  the time evolution of the measure $\mu_t(x)$ encodes the growth rule.
We refer to (\ref{eq:LDmu}) as
``Loewner domains'' in contrast to the ``Loewner chains'' discussed in Sec.~\ref{sec:2}. Although Loewner domains can be regarded as equivalent to Loewner chains---indeed, the densities  $\rho_t(x)$  and $\kappa_t(x)$ are obviously related to one another---, the former description seems to be more convenient for the study of growing interfaces (which encircle domains  with nonzero area).
%, while Loewner chains are more useful for describing increasing hulls defined by growing curves (chains).
From a more practical viewpoint, however, the difficult task that remains in the case of Loewner domains is to find the appropriate measure for physically relevant growth problems. In this context, an interesting open question concerns the possibility of generating rough (i.e., fractal) interfaces by means of a  Loewner evolution driven  by random measures, in analogy with the stochastic Loewner equation where random curves are produced \cite{review2,BB}. 
%These interesting problems certainly deserve further investigation.

\section{Summary and Conclusions}
\label{sec:4}

We have considered the problem of two-dimensional Laplacian growth in the context of the Loewner evolutions. We started by revisiting the problem of fingered growth and reported a novel exact solution of the chordal Loewner equation for the case of a symmetrical three-finger configuration. We then discussed a more general class of growth models in which  an interface grows from the real axis  into the upper-half plane.
Assuming that  the growth rule is specified in terms of a polygonal domain in the complex-potential plane and making appropriate use of the Schwarz-Christoffel transformation, we derived the corresponding  Loewner equation for the problem. Several examples were explicitly discussed, such as the case of a single ``extended finger,''  an interface with two tips, and the case of two competing interfaces. Although our model does not allow for a direct comparison with experiments, we have argued that its dynamics  is reminiscent of certain typical behaviors seen in Laplacian growth.  We have also briefly discussed a possible  extension of our Loewner equation to include the case in which the growth rule is specified in terms of a generic (not necessarily polygonal) curve. More generally, we have introduced the notion of ``Loewner domains'' in which the growth rule is described in terms of a time evolving signed measure. This opens up the possibility for studying   other interesting (and difficult) problems, such as the 
growth of fractal interfaces, within the context of stochastic Loewner evolutions. The extension to the radial geometry of the class  of growth models discussed here     is  another interesting open problem.

\begin{acknowledgments}
This work was supported in part by the Brazilian agencies FINEP, CNPq, and FACEPE and 
by the special programs PRONEX and CTPETRO. 

\end{acknowledgments}

%\begin{acknowledgements}
%\end{acknowledgements}


\begin{thebibliography}{}
\bibitem{loewner} K.~L\"owner, Math.~Ann.~{\bf 89}, 103 (1923).

\bibitem{univalent} P.~L.~Duren, {\it Univalent Functions} (Springer, New York, 1983).

\bibitem{review1} I.~A.~Gruzberg and L.~P.~Kadanoff, J.~Stat.~Phys.~{\bf 114}, 1183 (2004). 

\bibitem{review2} W.~Kager and B.~Nienhuis, J.~Stat.~Phys.~{\bf 115}, 1149 (2004).

\bibitem{BB} M.~Bauer and D.~Bernard,  {Phys.~Rep.} {\bf 432}, 115 (2006). 

\bibitem{review3} G.~Lawler, ``Introduction to the stochastic Loewner evolution," in V.~A.~Kaimanovich,
% K.~Schmidt, and W.~Woess, 
{\it Random walks and geometry} (Walter de Gruyter, Berlin,  2004), pp. 261--294.

\bibitem{poloneses} T.~Gubiec and P.~Szymczak, Phys.~Rev.~ E {\bf 77}, 041602 (2008).

\bibitem{schramm} O.~Schramm, Israel J.~Math.~{\bf 118}, 221 (2000).

\bibitem{selander} G.~Selander, Ph.D. thesis, Royal Institute of Technology, Stockholm, Sweden, 1999 (unpublished).

\bibitem{makarov} L.~Carleson and N.~Makarov, J.~Anal.~Math.~{\bf 87}, 103 (2002). 

\bibitem{combustion} O.~Zik and E.~Moses, Phys.~Rev.~E {\bf 60}, 518 (1999).

\bibitem{pelce} P.~Pelc\'e, {\it Dynamics of curved fronts}  (Academic Press, San Diego, 1988).

\bibitem{stanley} A.~L. ~Barab\'asi and H.~E.~Stanley, {\it Fractal concepts in surface growth} (Cambridge University Press, Cambridge, 1995). 

%\bibitem{hastings_levitov} M.~B.~Hastings and L.~S.~Levitov, Physica D {\bf 116}, 244 (1998).

%\bibitem{procaccia} B.~Davidovitch, H.~G.~E.~Hentschel, Z.~Olami, I.~Procaccia, L.~M.~Sander, and E.~Somfai, Phys.~Rev.~E {\bf 59}, 1368 (1999);
%B.~Davidovitch, M.~J.~Feigenbaum, H.~G.~E.~Hentschel, and I.~Procaccia, Phys.~Rev.~E {\bf 62}, 1706 (2000); F.~Barra, B.~Davidovitch, and I.~Procaccia 
%Phys.~Rev.~E {\bf 65}, 046144 (2002).	

%\bibitem{hastings} M.~B.~Hastings, Phys.~Rev.~E {\bf 64}, 046104 (2001).

%\bibitem{HS} D.~Bensimon, L.~P.~Kadanoff, S.~Liang, B.~I.~Shraiman, and C.~Tang, Rev.~Mod.~Phys.~{\bf 58}, 977 (1986).


\bibitem{kadanoff2} W.~Kager, B.~Nienhuis, and L.~P.~Kadanoff, J.~Stat.~Phys.~{\bf 115}, 805 (2004).


\bibitem{us} M.~A.~Dur\'an and G.~L.~Vasconcelos (unpublished).

\bibitem{CKP} G.~F.~Carrier, M.~Krook, and C.~E.~Pearson, {\it Functions of a complex variable: theory and technique} (Hod Books, Ithaca, 1983).


\bibitem{kennedy2} T.~Kennedy, J.~Stat.~Phys.~{\bf 137}, 839 (2009). 


\end{thebibliography}


\end{document}